\documentclass[pra,onecolumn,nofootinbib,12pt]{revtex4}

\usepackage{float}
\usepackage{graphicx}
\usepackage{array}
\usepackage{delarray}
\usepackage{amsmath}
\usepackage{amssymb}

\newcommand{\yb}{Yb${}^+\:$}

\begin{document}
\title{Laser cooling of trapped ytterbium ions with an ultraviolet diode laser}

\author{D. Kielpinski and M. Cetina}

\affiliation{Center for Ultracold Atoms and Research Laboratory of Electronics, Massachusetts Institute of Technology,
Cambridge MA 02139}

\author{J.A. Cox and F.X. K\"artner}
\affiliation{Optics and Quantum Electronics Group, Research Laboratory of Electronics, Massachusetts Institute of Technology,
Cambridge MA 02139}

\begin{abstract}
We demonstrate an ultraviolet diode laser system for cooling of trapped ytterbium ions. The laser power and linewidth are comparable to previous systems based on resonant frequency doubling, but the system is simpler, more robust, and less expensive. We use the laser system to cool small numbers of ytterbium ions confined in a linear Paul trap. From the observed spectra, we deduce final temperatures $< 270$ mK.
\end{abstract}

\maketitle

Laser-cooled trapped ions are the basis for many emerging technologies in atomic physics, especially large-scale ion-trap quantum computing \cite{Wineland-Meekhof-expt-issues-ion-QC,Kielpinski-Wineland-QCCD} and optical frequency metrology \cite{Diddams-Wineland-HgII-clock,Gill-optical-clock-rev}. In particular, laser-cooled \yb has been extensively investigated for optical clocks, owing to the narrow linewidths associated with transitions from the ground state to a number of metastable states \cite{Stenger-Telle-YbII-StoD3_2-clock,Hosaka-Klein-YbII-octupole-freq-std}. However, laser cooling of ions is often impeded by the expense and fragility of the required single-frequency ultraviolet (UV) laser source, which generally consists of a single-frequency infrared laser whose output is then frequency-doubled in an external optical resonator. This approach is currently adopted for laser cooling of \yb \cite{Bell-Klein-YbII-935-repump,Gill-Taylor-YbII-D_5/2-frequency,Sugiyama-single-YbII-trap,Stenger-Telle-YbII-StoD3_2-clock,Hosaka-Klein-YbII-octupole-freq-std}.\\ \\

Here we describe a simpler, more robust, and less expensive laser system for cooling \yb based on a UV laser diode. The laser diode is tuned to the \yb resonance wavelength at 369.5 nm using a combination of temperature tuning and optical feedback from a diffraction grating. Using this system, we have achieved Doppler cooling of small numbers of trapped \yb ions to a temperature of $< 270$ mK. The output power and linewidth of the laser are comparable to laser systems based on resonantly doubled Ti:sapphire lasers \cite{Bell-Klein-YbII-935-repump,Sugiyama-single-YbII-trap}. The simplicity of the laser system makes \yb an easily accessible and attractive candidate for future work with trapped ions. To the best of our knowledge, no shorter-wavelength laser has ever been used for laser cooling.\\ \\

A schematic of the UV laser system is shown in Fig. 1. The laser system was based on a commercial GaAlN laser diode (Nichia model NDHU110APAE2). We deposited an antireflection coating of $\mbox{Al}_2\mbox{O}_3$ onto the diode facet by electron-beam evaporation, monitoring the threshold current {\it in situ} to minimize the facet reflectivity. We stopped the coating when the threshold current exceeded 80 mA so as not to damage the device. Since the threshold current of the device before coating was 44.8 mA, we deduce a residual facet reflectivity of a few parts in a thousand \cite{Hildebrandt-Schael-AR-coated-blue-diode}. After antireflection coating, we inserted the diode into a Littrow-type external cavity \cite{Tino-Barsanti-Littrow-ECDL} formed with a 2400 line $\mbox{mm}^{-1}$ diffraction grating. Approximately 70\% of the optical power was fed back into the diode, with the long axis of the diode emission pattern parallel to the grating k-vector. Two stages of thermoelectric cooling maintained the laser diode at $-23 {}^{\circ}$C and the grating cavity at $-10 {}^{\circ}$C. The angular resolution of the grating mirror mount was insufficient for reliable wavelength tuning at the few pm level; we inserted a piezoelectric transducer (PZT) behind the tuning screw for fine adjustment. The laser threshold current was 35.5 mA for an output wavelength of 369.5 nm. Typically we operated the laser at 45 mA to get an output power of $400 \:\mu$W.\\ \\

\begin{figure}[htb]
\centerline{\includegraphics[width=8.3cm]{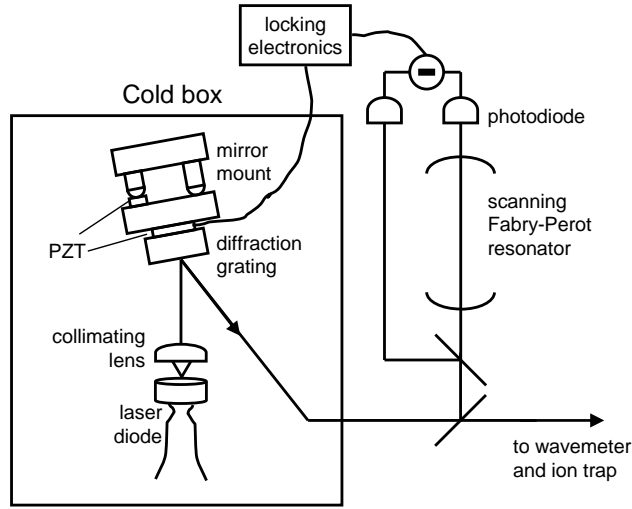}}
 \caption{Schematic of the UV laser system. A diffraction grating provides optical feedback to the UV diode in Littrow configuration. The laser cavity sits in a sealed cold box. The laser frequency is monitored and controlled by observing transmission through a scanning Fabry-Perot resonator. PZT, piezoelectric transducer.}
\end{figure}

We used a confocal scanning Fabry-Perot resonator to characterize the laser spectrum. The laser operated on a single frequency for periods exceeding one hour. Adjusting the current by $\sim 0.1$ mA was sufficient to restore single-frequency behavior at the original wavelength after a mode hop. We believe the mode hopping was driven by slow drifts of the external cavity temperature and could probably be eliminated with better thermal isolation. We controlled the laser frequency by translating the grating with a second PZT and synchronously varying the laser current \cite{Ricci-Haensch-ECDL-design}, obtaining mode-hop-free scan ranges of 3 -- 4 GHz. During single-frequency operation, we actively stabilized the laser frequency to a Fabry-Perot resonance using the sidelock technique \cite{Helmcke-Hall-sidelock}. From the in-loop photodetector noise while the laser was locked, we estimate the laser frequency stability to be better than 1 MHz integrated over a detection band of 1 Hz to 1 MHz. On longer timescales, the laser frequency drifted by as much as 30 MHz per minute, as estimated from the laser-cooling spectra (see below).\\ \\

The ion trap used to demonstrate laser cooling was a linear Paul trap \cite{Raizen-Wineland-linear-trap}. The electrode structure consisted of four parallel molybdenum rods with their axial centers on a square 6.4 mm on a side, along with two molybdenum tubes of inner diameter 2.25 mm parallel to the rods with their centers on the center of the square and separated axially by 50 mm. Transverse pseudopotential confinement was generated by applying radiofrequency voltage of frequency 1.9 MHz and amplitude 300--500 V peak-to-peak across pairs of diagonally opposite rod electrodes. Charging the tube electrodes to 10 V provided axial confinement. We loaded a few hundred ions into the trap by electron-impact ionization of ytterbium vapor with naturally occurring isotope abundances.\\ \\

We cooled the ${}^{172}\mbox{Yb}^+$ portion of the ions (natural abundance 22\%) by using the ultraviolet diode laser light to drive the $\mbox{S}_{1/2}$ -- $\mbox{P}_{1/2}$ transition near 369.5 nm. This transition is not quite closed; the $\mbox{P}_{1/2}$ state decays to the metastable $\mbox{D}_{3/2}$ state with branching ratio 0.048 \cite{Yu-Maleki-YbII-D-lifetimes}. To avoid the resulting interruption of laser cooling, we used another external-cavity diode laser near 935.2 nm to drive the $\mbox{D}_{3/2}$ -- ${}^3[3/2]_{1/2}$ ``repump'' transition. The ${}^3[3/2]_{1/2}$ state primarily decays to the ground $\mbox{S}_{1/2}$ state, returning the ion to the cooling cycle. The repump laser was sidelocked to a second Fabry-Perot resonator for short-term stability. We modulated the repump laser current at 900 kHz, well above the locking bandwidth, to minimize the effect of repump laser frequency drifts. The UV and repump lasers were focused weakly through the trap with $1/e^2$ beam waists of 0.36(3) mm. The UV power was 340(20) $\mu$W inside the vacuum chamber; the repump power was 4 mW. The UV and the repump lasers each saturated their respective transitions.\\ \\

We scanned the UV laser over the $\mbox{S}_{1/2}$ -- $\mbox{P}_{1/2}$ resonance by locking it to the Fabry-Perot and scanning the Fabry-Perot length with a PZT. In each instance the scan started from a red detuning of several hundred MHz and proceeded over the resonance at rates between 10 to 100 MHz ${\mbox s}^{-1}$; there was no dependence of line shape or width on scan rate. Ion fluorescence perpendicular to the trap axis was collimated with an f/1.4 plano-convex lens and refocused onto a photomultiplier tube (PMT). Stray light was blocked with a 370 nm interference filter. A sensitive transimpedance amplifier integrated the PMT pulses over a few tens of ms, generating a voltage proportional to light intensity.\\ \\

A typical spectrum of cold ions is shown in Fig. 2. As the laser scanned to the blue of resonance, the ions heated rapidly and the ion fluorescence fell dramatically. To quantify the homogeneous and inhomogeneous broadening in this unusual situation, we fit the spectrum to a Voigt profile multiplied by a soft step function of arctangent profile. Over a large number of scans, the best-fit Voigt profile displayed pure Lorentzian character with linewidth $45 \pm 3$ MHz, as compared to the natural linewidth of 19.8 MHz \cite{Pinnington-Kernahan-YbII-lifetimes}. The step function width of $1.5 \pm 0.3$ MHz consistently reproduced the laser linewidth estimated above, demonstrating the robustness of our fitting procedure. The power broadening contribution to the linewidth is predicted to be 42(5) MHz at the measured UV intensity of 170(40) mW $\mbox{cm}^{-2}$, in good agreement with the data. We estimate the laser frequency drift from drift in the apparent line center, which amounted to $\pm 0.5 \:\mbox{MHz} \:\mbox{s}^{-1}$ and is likely due to temperature drifts of the Fabry-Perot locking cavity. Since the laser scanned over the line in 2--4 s, laser frequency drift accounts for most of the variation in measured linewidth. \\ \\

\begin{figure}[htb]
\centerline{\includegraphics[width=8.3cm]{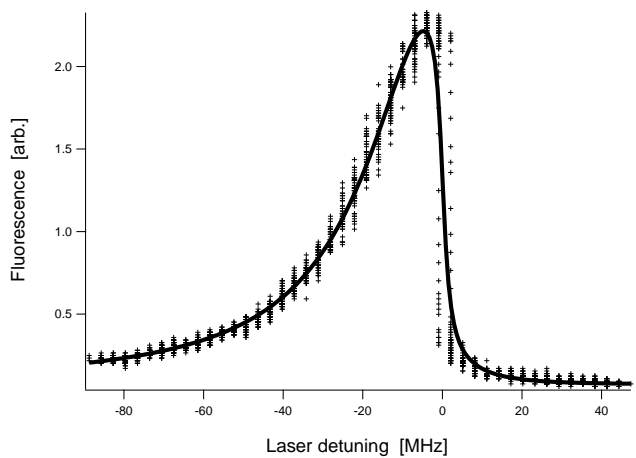}}
 \caption{Typical spectrum of cold ions. Crosses: data. Solid line: best fit to modified Voigt profile. Total linewidth is 42 MHz, mostly due to power broadening. The residual linewidth is $< 23$ MHz, corresponding to a temperature $< 270$ mK. See text for details.}
\end{figure}

The temperature $T$ of the trapped ions can be estimated from the Gaussian (Doppler broadening) contribution to the Voigt linewidth by the formula
\begin{equation}
T = \frac{m c^2}{8 k_B \ln 2} \frac{\Delta \nu_G}{\nu}
\end{equation}
\noindent where $m$ is the mass of ${}^{172}\mbox{Yb}^+$, $\Delta \nu_G$ is the Gaussian width (FWHM), and $\nu$ is the transition frequency. Since the lineshape is essentially Lorentzian, we set an upper limit on $\Delta \nu_G$ by considering the upper limit (48 MHz) of the observed linewidths and deconvolving with the lower limit (37 MHz) of the predicted power broadening. Using the Voigt deconvolution formula $\Delta \nu_D = \sqrt{\Delta \nu_{\rm tot} (\Delta \nu_{\rm tot} - \Delta \nu_L)}$, we find the upper limit $\Delta \nu_G < 23$ MHz. The corresponding upper limit on the temperature is $T < 270$ mK, comparable to results obtained using resonantly doubled laser systems \cite{Gill-Taylor-YbII-D_5/2-frequency,Sugiyama-single-YbII-trap}.\\ \\

We have demonstrated a UV diode laser system for cooling of \yb ions. Our laser system provides single-frequency light with roughly the same power (400 $\mu$W) and linewidth ($\sim 1$ MHz) as traditional systems based on resonant frequency doubling. Our system is much simpler, more robust, and less expensive than these past systems. We have used the laser system to cool \yb ions in a linear Paul trap to temperatures $< 270$ mK, demonstrating that our system is a viable choice for Doppler cooling of \yb. Future improvements include locking the UV laser to a long-term frequency reference and injection-locking a second UV diode laser for increased power.\\ \\

We thank Peter O'Brien of MIT Lincoln Laboratory for assistance with antireflection coating of the laser diodes. We also wish to acknowledge helpful discussions and assistance from Vladan Vuleti\'c and Isaac Chuang. This work was supported by the US Air Force Office of Scientific Research under contract F49620-03-1-0313. D.K. was also supported by a Pappalardo Fellowship. D.K.'s email address is d.kielpinski@griffith.edu.au.\\ \\


\begin{thebibliography}{99}
\bibitem{Wineland-Meekhof-expt-issues-ion-QC} D.J. Wineland, C. Monroe, W.M. Itano, D. Leibfried, B.E. King, and D.M. Meekhof, J. Res. NIST 103, 259 (1998).
\bibitem{Kielpinski-Wineland-QCCD} D. Kielpinski, C. Monroe, and D.J. Wineland, Nature 417, 709 (2002).
\bibitem{Diddams-Wineland-HgII-clock} S.A. Diddams, Th. Udem, J.C. Bergquist, E.A. Curtis, R.E. Drullinger, L. Hollberg, W.M. Itano, W.D. Lee, C.W. Oates, K.R. Vogel, and D.J. Wineland, Science 293, 825 (2001).
\bibitem{Gill-optical-clock-rev} P. Gill, Metrologia 42, S125 (2005).
\bibitem{Stenger-Telle-YbII-StoD3_2-clock} J. Stenger, C. Tamm, N. Haverkamp, S. Weyers, and H.R. Telle, Opt. Lett. 26, 1589 (2001).
\bibitem{Hosaka-Klein-YbII-octupole-freq-std} K. Hosaka, S.A. Webster, P.J. Blythe, A. Stannard, D. Beaton, H.S. Margolis, S.N. Lea, and P. Gill, IEEE Trans. Inst. Meas. 54, 759 (2005).
\bibitem{Bell-Klein-YbII-935-repump} A.S. Bell, P. Gill, H.A. Klein, A.P. Levick, C. Tamm, and D. Schnier, Phys. Rev. A 44, 20 (1991).
\bibitem{Gill-Taylor-YbII-D_5/2-frequency} P. Gill, H.A. Klein, A.P. Levick, M. Roberts, W.R.C. Rowley, and P. Taylor, Phys. Rev. A 52, R909 (1995).
\bibitem{Sugiyama-single-YbII-trap} K. Sugiyama, Jpn. J. Appl. Phys 38, 2141 (1999).
\bibitem{Hildebrandt-Schael-AR-coated-blue-diode} L. Hildebrandt, R. Knispel, S. Stry, J.R. Sacher, and F. Schael, Appl. Opt. 42, 2110 (2003).
\bibitem{Tino-Barsanti-Littrow-ECDL} G.M. Tino, L. Hollberg, A. Sasso, M. Inguscio, and M. Barsanti, Phys. Rev. Lett. 64, 2999 (1990).
\bibitem{Ricci-Haensch-ECDL-design} L. Ricci, M. Weidem\"uller, T. Esslinger, A. Hemmerich, C. Zimmermann,
V. Vuletic, W. K\"onig, and T.W. H\"ansch, Opt. Comm. 117, 541 (1995).
\bibitem{Helmcke-Hall-sidelock} J. Helmcke, S.A. Lee, and J.L. Hall, Appl. Opt. 21, 1686 (1982).

\bibitem{Raizen-Wineland-linear-trap} M.G. Raizen, J.M. Gilligan, J.C. Bergquist, W.M. Itano, and D.J. Wineland, J. Mod. Opt 39, 233 (1992).
\bibitem{Yu-Maleki-YbII-D-lifetimes} N. Yu and L. Maleki, Phys. Rev. A 61, 022507 (2000).
\bibitem{Pinnington-Kernahan-YbII-lifetimes} E. H. Pinnington, G. Rieger, and J. A. Kernahan, Phys. Rev. A 56, 2421 (1997).


\end{thebibliography}
\end{document}